\documentclass[9pt,conference]{IEEEtran}
\IEEEoverridecommandlockouts

\usepackage{cite}
\usepackage{amsmath,amssymb,amsfonts}
\usepackage{algorithmic}
\usepackage{graphicx}
\usepackage{textcomp}
\usepackage{xcolor}
\usepackage{booktabs}
\usepackage{colortbl}
\usepackage{arydshln}
\usepackage{caption}
\usepackage{hyperref}
\def\BibTeX{{\rm B\kern-.05em{\sc i\kern-.025em b}\kern-.08em
    T\kern-.1667em\lower.7ex\hbox{E}\kern-.125emX}}
\begin{document}

\title{Ideal-LLM: Integrating Dual Encoders and Language-Adapted LLM for Multilingual Speech-to-Text}

\author{\IEEEauthorblockN{Hongfei Xue$^*$\thanks{$^*$Equal contribution.}}
\IEEEauthorblockA{
\textit{Northwestern Polytechnical University}\\
hfxue@mail.nwpu.edu.cn}
\and
\IEEEauthorblockN{Wei Ren$^*$}
\IEEEauthorblockA{
\textit{Chongqing Changan Automobile Co., Ltd}\\
agnanren@tinnove.com.cn}
\and
\IEEEauthorblockN{Xuelong Geng}
\IEEEauthorblockA{
\textit{Northwestern Polytechnical University}\\
xlgeng@mail.nwpu.edu.cn}
\and
\IEEEauthorblockN{Kun Wei}
\IEEEauthorblockA{
\textit{Northwestern Polytechnical University}\\
ethanwei@mail.nwpu.edu.cn}
\and
\IEEEauthorblockN{Longhao Li}
\IEEEauthorblockA{
\textit{Northwestern Polytechnical University}\\
lhli@mail.nwpu.edu.cn}
\and
\IEEEauthorblockN{Qijie Shao}
\IEEEauthorblockA{
\textit{Northwestern Polytechnical University}\\
qjshao@npu-aslp.org}
\and
\IEEEauthorblockN{Linju Yang}
\IEEEauthorblockA{
\textit{Chongqing Changan Automobile Co., Ltd}\\
louisyang@tinnove.com.cn}
\and
\IEEEauthorblockN{Kai Diao}
\IEEEauthorblockA{
\textit{Chongqing Changan Automobile Co., Ltd}\\
diaokai@changan.com.cn  }
\and
\IEEEauthorblockN{Lei Xie$^{\dagger}$\thanks{$^\dagger$Corresponding author.}}
\IEEEauthorblockA{\
\textit{Northwestern Polytechnical University}\\
lxie@nwpu.edu.cn}
}


\maketitle

\begin{abstract}
Integrating audio encoders with LLMs through connectors has enabled these models to process and comprehend audio modalities, significantly enhancing speech-to-text tasks, including automatic speech recognition (ASR) and automatic speech translation (AST). However, these methods often overlook the critical aspect of language adaptation in multilingual settings, relying instead on multilingual data without adequately addressing language differences. To address this gap, we propose the Ideal-LLM model, which employs dual multilingual encoders to enrich language feature information and utilizes a language-adapted connector to target the adaptation of each language specifically. By leveraging the complementary strengths of Whisper and MMS encoders, our approach ensures richer multilingual representations. Additionally, the language-adapted connector enhances modal transformation via a language weight selector tailored for each language. Experimental results demonstrate that Ideal-LLM significantly improves ASR performance, achieving a 32.6\% relative reduction in average word error rates compared to the standard speech encoder integrated with LLMs and yields an average BLEU score of 36.78 for AST task.
\end{abstract}


\section{Introduction}
\label{sec:intro}
Text-based Large Language Models (LLM) have demonstrated significant influence in the field of artificial intelligence due to their powerful natural language understanding and generation capabilities~\cite{openai2022chatgpt, openai2023gpt4, brown2020language, anil2023palm, LLaMA}. Recently, researchers have explored integrating audio encoders with LLMs through connectors, enabling LLMs to process and understand audio modalities~\cite{gong2023listentu, tang2023salmonn, chu2023qwenaudio, 24wavllm, chu2024qwen2audio, 24echat}. By training with audio-text paired data, the connector aligns the output space of the audio encoder with the input space of the LLM, playing a crucial role in audio understanding tasks.

Multilingual speech-to-text (S2T) is a vital task in audio understanding, encompassing both multilingual automatic speech recognition (ASR) and automatic speech translation (AST). Existing studies have shown that integrating speech encoders and LLMs with a connector significantly enhances performance compared to traditional end-to-end models~\cite{meta24llmasr, 23speechllama, chu2024qwen2audio, 24slam-asr, bai2024seed, 24s2tt}. For example, one approach \cite{meta24llmasr} employs a connectionist temporal classification (CTC)~\cite{20ctc} trained encoder to process speech sequences, which are then fed into an LLM decoder through a projection layer. This method outperforms end-to-end models in multilingual ASR tasks. Similarly, Speech-LLaMA~\cite{23speechllama} uses a CTC compressor and a simple audio encoder to map compressed acoustic features into the continuous semantic space of the LLM, achieving superior performance on several AST test sets.
Moreover, approaches leveraging Whisper encoders~\cite{23whisper} or self-supervised learning (SSL) encoders have demonstrated superior improvements~\cite{chu2023qwenaudio, 24slam-asr, bai2024seed, 24s2tt}. Qwen-Audio~\cite{chu2023qwenaudio} and Qwen2-Audio~\cite{chu2024qwen2audio} leverage a fine-tuned Whisper encoder~\cite{23whisper} to extract speech representations, leading to significant advancements in both multilingual ASR and AST tasks. Seed-ASR~\cite{bai2024seed} has achieved state-of-the-art (SOTA) results in multilingual, multi-domain ASR test sets by feeding continuous speech representations and contextual information into the LLM, fully exploiting the LLM's capabilities.
Additionally, a recent study\cite{24s2tt} employs W2v-BERT~\cite{21w2vbert} as a speech encoder, inputs it into the LLM via a length adapter, and incorporates Chain of Thought (COT) to achieve SOTA results on the CommonVoice~\cite{19commonvoice} and CovoST2~\cite{21covost2} test sets for AST.

Despite these advancements, existing studies often focus on adding multilingual data rather than adequately considering language adaptation. In terms of connectors, the study in~\cite{23connectmodels} delves into various connector architectures, revealing that Q-former~\cite{li2023blip} surpasses both linear layers and multi-head cross-attention mechanisms~\cite{17attention}. Similarly, the study in~\cite{geng2024unveiling} employs HuBERT~\cite{21hubert} to extract speech representations and utilizes a Transformer encoder~\cite{17attention} as a connector to map these representations into the LLM's space, achieving excellent results on several Mandarin test sets. However, these connectors primarily rely on decoder loss for optimization and lack language adaptation, potentially limiting their effectiveness in accurately mapping the multilingual representation space.

\begin{figure*}[ht]
\centering
\includegraphics[width=0.80\linewidth]{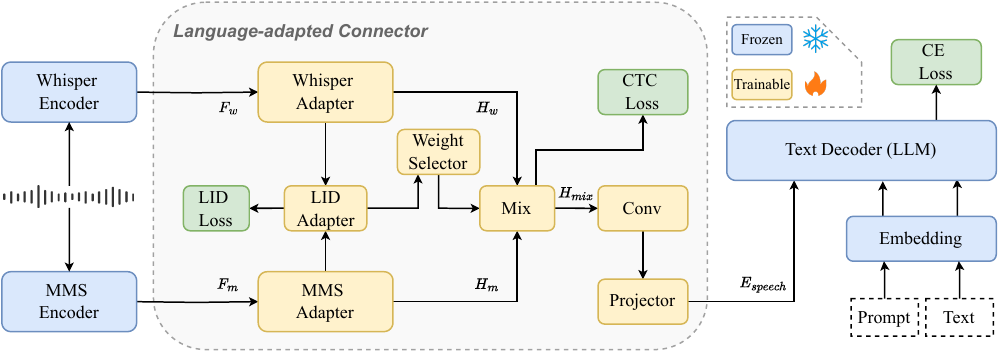}
\caption{
The overall framework of the proposed Ideal-LLM model.}
\label{fig:model}
\vspace{-5pt}
\end{figure*}

While existing models have leveraged the powerful text comprehension capabilities of LLMs to enhance multilingual S2T tasks, aligning the speech feature space of each language to the LLM remains insufficient due to inherent linguistic differences. On one hand, the features extracted by the encoder may not be sufficiently adapted for multiple languages. On the other hand, existing connectors cannot specifically align two representation spaces for every language.
To address this problem, we propose the Ideal-LLM model, which employs dual multilingual encoders to enrich the language information contained in speech features and utilizes a language-adapted connector to target the adaptation of each language specifically. We leverage Whisper~\cite{23whisper} and MMS~\cite{23mms}, two popular and robust models trained on extensive multilingual data using weakly-supervised and self-supervised learning, respectively. These models' representations complement each other due to their distinct pre-training methods on various language distributions~\cite{23comparison}. While Salmonn~\cite{tang2023salmonn} and WavLLM~\cite{24wavllm} also employ dual encoders, they focus on extracting semantic and acoustic information separately rather than addressing the critical aspect of language adaptation. Additionally, our language-adapted connector facilitates modal alignment through CTC loss and integrates dual encoder representations via a language weight selector. The experimental results indicate that our model is more effective at distinguishing languages and aligning the multilingual embedding space. Specifically, our approach significantly enhances ASR performance, achieving a 32.6\% relative reduction in average word error rates (WER) compared to Whisper encoder integrated with LLMs. In AST task, our method yields an average BLEU score of 36.78, surpassing the performance of Qwen2-Audio~\cite{chu2024qwen2audio}.
\vspace{-7pt}

\section{Method}
\vspace{-7pt}
\label{sec:proposed_method}
\subsection{Architecture}
\vspace{-1pt}
Our model comprises dual encoders, a language-adapted connector, and a text decoder. An illustration of the overall architecture is shown in Fig~\ref{fig:model}.

\textbf{Dual Encoders} 
Our dual speech encoders are based on Whisper~\cite{23whisper} and MMS~\cite{23mms}, which are robust models trained on large multilingual datasets using weakly-supervised and self-supervised learning, respectively. We process the input speech through two paths. First, we convert the speech signal to an 80-channel log-magnitude Mel spectrogram representation, which is input to the Whisper encoder to obtain the speech feature $F_w$. Second, the speech signal is directly input to the MMS encoder to obtain the speech feature $F_m$. 

\textbf{Language-adapted Connector} 
Since the dual encoders have been trained on different distributions of language data, we design a connector to perform a language-dependent fusion of the dual encoders' features and transform them into the embedding space of the LLM. First, the speech features $F_w$ and $F_m$ are transformed into hidden representations by Whisper and MMS adapter, which are transformer encoder networks~\cite{17attention}, resulting in $H_w$ and $H_m$, respectively. Then, based on the Weight Selector, $H_w$ and $H_m$ are mixed with different weights to form $H_{mix}$. If the frame lengths differ, they are aligned by adding blank frames. Finally, the fused hidden representations $H_{mix}$ are sequentially downsampled through the convolutional layer, and the projection layer maps them to $E_{speech}$ in the LLM embedding space.

In the Weight Selector, we initialize trainable parameters for each language and apply a sigmoid function to generate weights. When the LID Adapter predicts a specific language, it selects the parameters of the specified language for weighting. This process is learned through backpropagation with decoder loss and LID loss, guiding the model to prefer certain encoders. The specific formula is as follows:
\begin{align}
l &= \text{LID Adapter}(L) ,\\
w &= \text{Weight Selector}(l, W) ,\\
w' &= \text{sigmoid}(w) ,\\
H_{mix} &= H_w \cdot (1 - w') + H_m \cdot w',
\end{align}
where the Weight Selector contains a set of learnable weights $W$ that are selected according to a specific language $l$ from the LID Adapter in $L$, corresponding to the weights $w$.

\textbf{Text Decoder} 
The text decoder is built upon the phi-3-mini model~\footnote{https://huggingface.co/microsoft/Phi-3-mini-4k-instruct}, a language model with 3.8 billion parameters trained on 3.3 trillion tokens~\cite{abdin2024phi}. This model demonstrates robust overall performance. The prompt and text labels are represented by the tokenizer embedding layer, which is then concatenated with $E_{speech}$ from the language-adapted connector. These embeddings are then fed into the text decoder, with the output target being the text labels.

\subsection{Multi-task Training}
Our multi-task training includes Cross-Entropy (CE) loss for the decoder, CTC loss and LID loss for the language-adapted connector. An illustration of the overall architecture is shown in Fig~\ref{fig:model}.

\textbf{CE Loss}
The CE loss is used in the LLM to optimize the model's final recognition or translation results.

\textbf{CTC Loss}
To facilitate the transformation of speech representations into the LLM's textual representations in the language-adapted connector, we employ an additional CTC loss in $H_{mix}$ to impose constraints.

\textbf{LID Loss}
We incorporate an LID prediction loss to select appropriate weights based on specific languages to enhance representation fusion. Specifically, the LID Adapter performs pooling operations to reduce $H_m$ and $H_w$ to one-dimensional representations, which are then summed to form LID representations. The LID loss optimizes these representations for language prediction. Once a specific language is predicted, the Weight Selector gets the corresponding weight, which is sent to the mixing component for weighted fusion.

\begin{table*}
\centering
\caption{WER (\%) results on Multilingual Librispeech for different methods. For the Baseline and Ideal-LLM Base models, we use 10 kh English, while all other models are 44 kh.}
\label{tab:MLSASR}
\resizebox{1.0\textwidth}{!}
{
\begin{tabular}{@{}cccccccccccc@{}}
\toprule
Model                                      & Training Step (k) & Trainable Params (B)     & \textit{en}              & \textit{de}               & \textit{nl}                & \textit{fr}               & \textit{es}              & \textit{it}              & \textit{pt}           & \textit{pl}               & Avg                  \\ \midrule
\multicolumn{1}{l}{Duration (kh)} & {}              & {} & 10 / 44 & 1.97  & 1.55 & 1.08 & 0.9 & 0.25 & 0.16 & 0.10 & {} \\ \midrule
\multicolumn{1}{l}{MMS CTC~\cite{23mms}}                                   & 50          & 1.0                   & -                    & -                    &  -                   &  -                   &  -                   &  -                   &  -                   &  -                   & 8.7                  \\
\multicolumn{1}{l}{LLaMA with ASR~\cite{meta24llmasr}}                                    & 250         & 0.240           & 6.2                  & \textbf{6.7}                  & 11.3                 & 5.5                  & 5.2                  & 10.8                 & 16.2                 & 15.9                 & 9.73                 \\ \midrule
\multicolumn{1}{l}{Baseline}                                          & 26          & 0.075                 & 7.42                 & 9.55                 & 14.05                & 8.18                 & 7.51                 & 14.64                & 12.69                & 14.60                & 11.59                \\
\multicolumn{1}{l}{Ideal-LLM Base}                              & 26          & 0.172                 & 7.44                 & 8.25                 & 12.47                & 6.71                 & 5.47                 & 11.84                & \textbf{10.87}                & 9.38                & 9.05                 \\
\multicolumn{1}{l}{Ideal-LLM Large}                              & 50          & 0.303                 & \textbf{6.15}                 & 7.12                 & \textbf{11.23}                & \textbf{5.40}                 & \textbf{4.26}                 & \textbf{9.93}                 & 12.41                & \textbf{6.02}                 & \textbf{7.81}                 \\ \bottomrule
\end{tabular}
}
\vspace{-5pt}
\end{table*}

The total loss is formulated as follows:
\begin{align}
L_{\text{all}} = (1 - \alpha) \cdot L_{\text{decoder}} + \alpha \cdot L_{\text{CTC}} + \beta \cdot L_{\text{LID}}
\end{align}
This training strategy ensures the model learns to perform recognition and translation while accurately weighting and fusing language-specific representations.
We will open source the training code once the paper is accepted.
\section{Experiments}
\label{sec:experiments}

\subsection{Datasets}
\textbf{Multilingual ASR}
For the multilingual ASR task, we use the Multilingual LibriSpeech (MLS) dataset~\cite{20mllibrispeech}, following prior work~\cite{meta24llmasr}. This dataset is a 50,000-hour ASR corpus derived from reading audiobooks on LibriVox. It comprises eight languages: English (\textit{en}), German (\textit{de}), Dutch (\textit{nl}), French (\textit{fr}), Spanish (\textit{es}), Italian (\textit{it}), Portuguese (\textit{pt}), and Polish (\textit{pl}). The dataset predominantly consists of English recordings, with 44,500 hours dedicated to this language. We evaluated performance using the WER metric.

\textbf{Multilingual AST}
For the multilingual AST task, we use the CovoST2 dataset~\cite{21covost2}, following Qwen2-Audio~\cite{chu2024qwen2audio}. To utilize the model obtained from the ASR task, we select the intersection of the languages in MLS: English to German (\textit{en-de}), German to English (\textit{de-en}), French to English (\textit{fr-en}), Spanish to English (\textit{es-en}), and Italian to English (\textit{it-en}). Performance is evaluated using BLEU scores, employing the sacrebleu tool for comparison~\footnote{https://github.com/mjpost/sacrebleu}.

\subsection{Experiment Setup}
\textbf{Baseline}
Ideal-LLM integrates dual encoders and a language-adapted LLM, making the most relevant baseline is a model that integrates an encoder with an LLM.
For the baseline model, we use a combination of the Whisper encoder and the phi-3-mini model, where the Whisper encoder is from Whisper Large-v3~\footnote{https://huggingface.co/openai/whisper-large-v3}. The connector consists of only the Whisper adapter, convolutional, and projector layers. The Whisper Adapter is a 4-layer Transformer encoder. The convolutional layer performs 2x downsampling, and the projector is a linear layer that maps feature dimensions to the LLM embedding dimensions. We use 10k hours of English data and the full data for the other seven languages during ASR task training for convenience. We also applied a data balancing strategy as in~\cite{21XLSR}. The Adam optimizer has a peak learning rate of 5e-4 and a warmup of 2k steps for 26k training steps. We use 8 NVIDIA 4090 GPUs with 24GB of memory each, with gradient accumulation equivalent to about 400s of data per GPU. During training, only the connector is trainable. For the AST task, training is initialized with parameters from the ASR model, using a peak learning rate of 1e-4, a warmup of 2k steps, and a total of 10k steps. The prompt for the ASR task is \textit{``Transcribe the speech to text,"} and for the AST task, it is \textit{``Translate the speech to \{language\}."}, where the language is English or German.

\textbf{Ideal-LLM Base}
In the proposed base model setup, the Whisper encoder remains the same as in the baseline model, and the MMS encoder is the 300M version~\footnote{https://huggingface.co/facebook/mms-300m}. Both Whisper and MMS adapters are 4-layer Transformer encoders. The LID Adapter is a linear layer that maps feature dimensions to 8 languages. During training, only the language-adapted connector is trainable. For multi-task training, we set $\alpha$ to 0.1 and $\beta$ to 0.05. The experimental data and training strategy are the same as the baseline model.

\textbf{Ideal-LLM Large}
For the large model, the MMS encoder uses the 1B version~\footnote{https://huggingface.co/facebook/mms-1b}. The Whisper Adapter and the MMS Adapter are 9-layer Transformer encoders, and the rest of the configurations are the same as for the Ideal-LLM base model. We use the full MLS dataset for training, which increases the amount of data mainly in English compared to the baseline and Ideal-LLM base models. The Adam optimizer has a peak learning rate of 2e-4, a warmup of 2000 steps, and 50k training steps. We use 8 NVIDIA A6000 GPUs with 48GB of memory each, with gradient accumulation equivalent to about 800s data per GPU. For inference, we use the best five models for average decoding. Training for the AST task is initialized with parameters from the ASR task model, using a peak learning rate of 5e-5, a warmup of 2k steps, and a total of 10k training steps.

\begin{table}[t]
\centering
\caption{BLEU scores (\%) for the AST task on the CoVoST2 dataset.}
\label{tab:covost2AST}
\resizebox{1.0\linewidth}{!}{
\begin{tabular}{lcccccc}
\toprule
                         & \textit{en-de} & \textit{de-en} & \textit{es-en} & \textit{fr-en} & \textit{it-en} & Avg   \\ \midrule
\multicolumn{1}{l}{Duration (h)} & 364  & 119 & 97 & 180  & 28  \\ \midrule
Speech-LLaMA~\cite{23speechllama}              &  -     & 27.1  & 27.9  & 25.2  & 25.9  &  -     \\
Qwen2-Audio~\cite{chu2024qwen2audio}              & \textbf{29.9}  & 35.2  & 40.0  & 38.5  & 36.3  & 35.98 \\ \midrule
Baseline                 & 22.7  & 34.1  & 38.6  & 36.1  & 34.7  & 33.24  \\
Ideal-LLM Base     & 23.3  & 34.4  & 39.3  & 37.1  & 34.4  & 33.70  \\
\ \ \ \ + COT & 24.6  & 37.8  & 40.9  & 38.7  & 37.4  & 35.88 \\
Ideal-LLM Large     & 23.7  & 34.9  & 39.9  & 37.8  & 35.2  & 34.30 \\
\ \ \ \ + COT & 25.9  & \textbf{38.5}  & \textbf{41.5}  & \textbf{40.0}  & \textbf{38.0}  & \textbf{36.78} \\ \bottomrule
\end{tabular}
}
\vspace{-10pt}
\end{table}

\subsection{Main Results}

\textbf{ASR Results}
Table~\ref{tab:MLSASR} presents the WER results for the ASR task on the MLS dataset, using two existing works for reference~\cite{23mms, meta24llmasr}. MMS CTC~\cite{23mms} uses the full MLS dataset with CTC training on a 1B MMS encoder and adds an n-gram language model for inference. LLaMA with ASR~\cite{meta24llmasr} employs an encoder trained with CTC loss on MLS data to generate speech features and uses LLaMA~\cite{LLaMA} as the text decoder with LoRA~\cite{22lora}.
The experimental results show that the Ideal-LLM Base significantly decreases WER across all languages compared to the baseline, achieving a relative decrease of 19.3\% in average WER. This indicates that our proposed model better utilizes multilingual speech representation. The Ideal-LLM large model, which includes more data, a larger encoder and an increased number of trainable parameters (0.13B), shows a relative decrease of 16.3\% in average WER, resulting in a 32.6\% decrease compared to the baseline model. Despite a slight regression in WER for Portuguese, we think this is due to the data imbalance caused by the addition of a large amount of English data and Portuguese is more sensitive to the imbalance. The Ideal-LLM large model also outperforms existing works, demonstrating its potential for multilingual ASR.

\begin{table}[t]
\centering
\caption{WER (\%) results on 8 languages of MLS in proposed base model for ablation study.}
\label{tab:ablantion}
\resizebox{0.85\linewidth}{!}{
\begin{tabular}{@{}lcc@{}}
\toprule
Model                             & Trainable params (B)  & Average   \\ \midrule
Ideal-LLM Base                     & 0.172         & \textbf{9.05}  \\
\ \ \ \ - Weight Selector           & 0.172          & 9.30  \\
\ \ \ \ \ \ \ \  - Dual Encoders & 0.116 & 10.52 \\
\ \ \ \ \ \ \ \ \ \ \ \ - CTC Loss   & 0.075   & 11.59 \\ \bottomrule
\end{tabular}}
\vspace{-10pt}
\end{table}

\textbf{AST Results}
Table~\ref{tab:covost2AST} shows the BLEU scores for the AST task on the CoVoST2 dataset~\cite{21covost2}, referencing two existing works~\cite{23speechllama, chu2024qwen2audio}. Speech-LLaMA~\cite{23speechllama} combines a simple encoder with LLaMA~\cite{LLaMA}, while Qwen2-Audio~\cite{chu2024qwen2audio} utilizes a fine-tuned Whisper~\cite{23whisper} and Qwen~\cite{bai2023qwen} decoder. Both baseline and Ideal-LLM models are configured as described in the previous section. We also employ a COT prompt to enhance performance. The prompt is: \textit{``First transcribe the speech to text, and then translate the speech to {language}."}, where {language} can be either English or German.
The experimental results reveal that both the Ideal-LLM base and large models increase BLEU scores compared to the baseline, with the large model achieving an increase of 1 point in average BLEU scores. The inclusion of the COT prompt further improves the BLEU scores of the Ideal-LLM models. Ideal-LLM Large with COT outperforms Qwen2-Audio, showcasing its potential for multilingual AST.

\vspace{-5pt}
\subsection{Analysis}

\textbf{Ablation Study}
We conduct ablation experiments on the dual encoder structure and the language-adapted connector proposed in the paper. The language-adapted connector comprises the CTC loss function for converting the speech feature to the text space and the Weight Selector module for learning language-specific weights.
From the experimental results shown in Table~\ref{tab:ablantion}, removing the Weight Selector structure increases the average WER from 9.05 to 9.30. Additionally, removing the dual-encoder structure leads to an 11.6\% relative increase in WER, highlighting the dual-encoder's critical role in providing richer multilingual representations. Furthermore, eliminating the CTC loss function results in a 9.2\% relative increase in WER, indicating that the inclusion of the CTC loss in the connector effectively aids the transformation of speech representation.

\begin{figure}[t]
\centering
\includegraphics[width=0.95\linewidth]{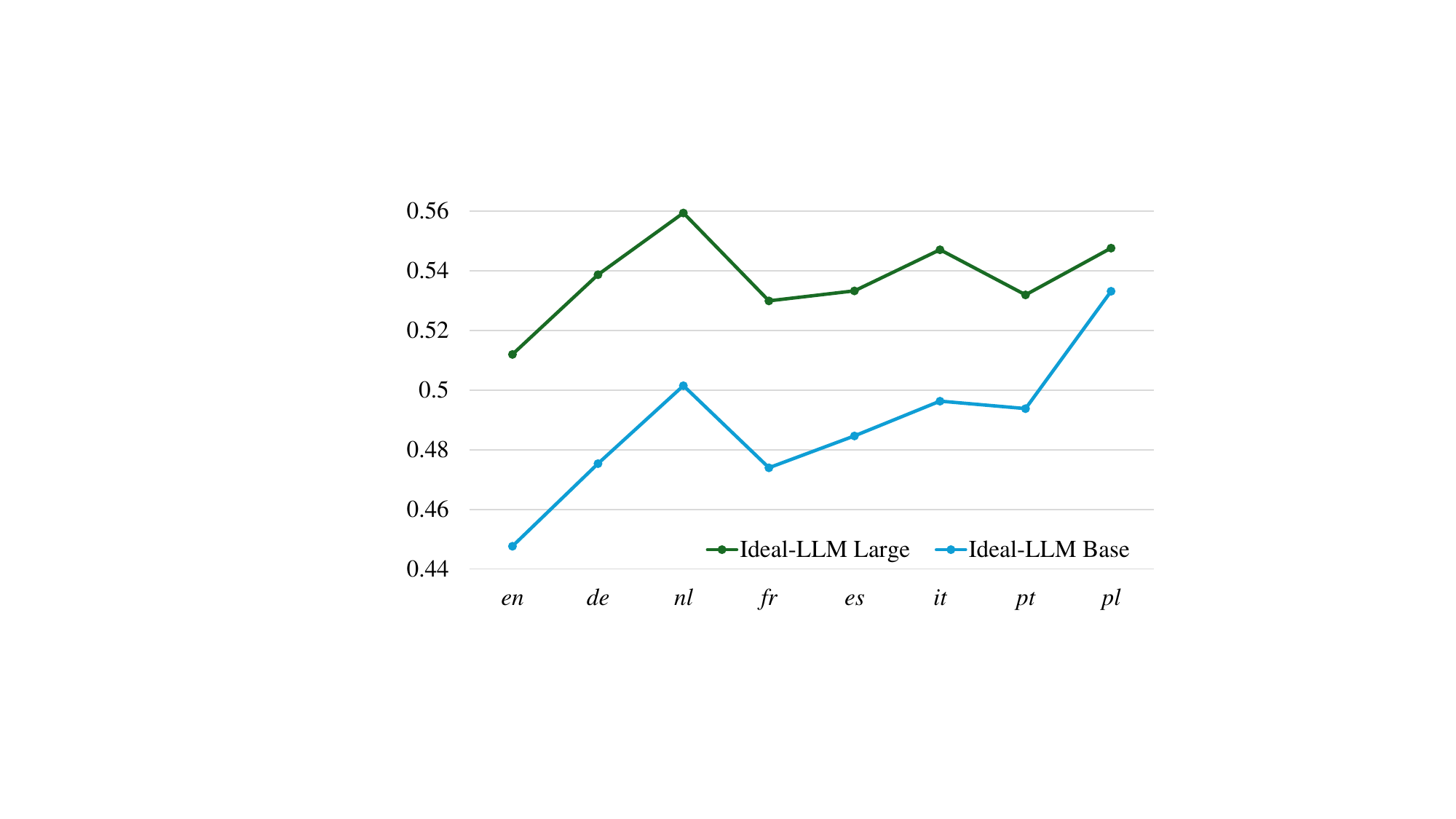}
\caption{Weight distribution of MMS Encoder ($w'$) in the Ideal-LLM base and large models across different languages.}
\label{fig:language}
\vspace{-5pt}
\end{figure}

\textbf{Weight Selector}
We analyze whether Weight Selector learns based on the pre-training methods and datasets of Whisper and MMS. As shown in Fig.~\ref{fig:language}, the blue and green lines represent the weights of the MMS encoder for the Ideal-LLM base and large models, respectively.
When comparing these two lines, we observe that the overall weight of Ideal-LLM large is approximately 0.06 higher than the base model. We attribute this to the stronger SSL capability of the MMS 1B encoder, demonstrating that the fusion is correlated with the pre-training methods. The overall trend for the two lines is similar, likely due to the same data distribution used during pre-training of MMS 1B and 300M.
The weight for \textit{en} reveals that since Whisper pre-trains with a significantly larger amount of \textit{en} data than MMS, it is weighted substantially lower than other languages. For languages such as \textit{nl}, \textit{it}, and \textit{pl}, Whisper is pre-trained with only 2k hours of data compared to approximately 20k for MMS, resulting in weights skewed in favor of MMS for these languages compared to the others.

\begin{figure}[t]
\centering
\includegraphics[width=1.0\linewidth]{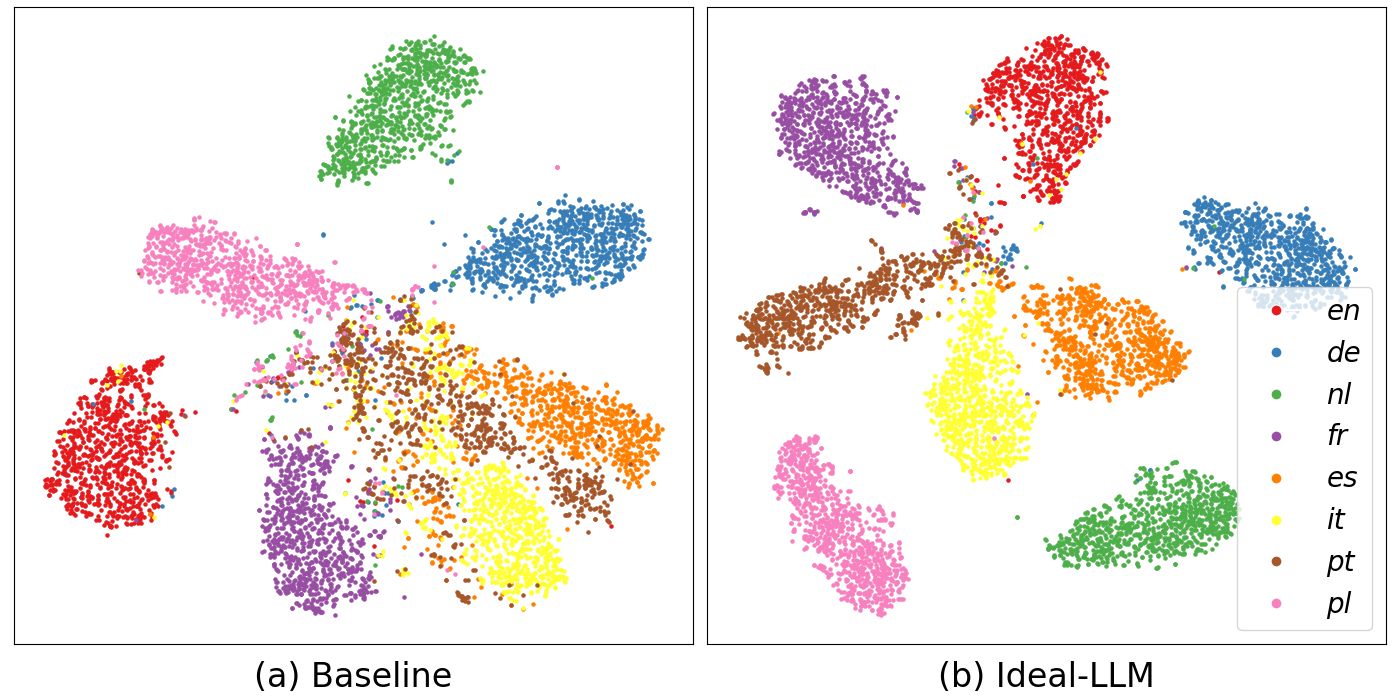}
\caption{T-SNE analysis for $E_{speech}$ of 900 utterances in each of the eight languages.}
\label{fig:distribution}
\vspace{-10pt}
\end{figure}

\textbf{Speech Embedding}
We conduct a T-SNE analysis of the speech embedding obtained from the Baseline model and the Ideal-LLM model. To verify the robustness of the models, we randomly selected 900 sentences for each language from the test set of CommonVoice~\cite{19commonvoice}, an out-of-domain dataset. As illustrated in Fig.~\ref{fig:distribution}, $E_{speech}$ from the Baseline model exhibits overlap among different languages, indicating its inability to effectively separate the languages during the conversion process. In contrast, the Ideal-LLM model demonstrates a marked reduction in overlap. This suggests that by utilizing a language-adapted connector, our proposed model successfully differentiates between languages during the embedding conversion process. This distinction is crucial for ensuring the accuracy and reliability of multilingual S2T tasks.

\textbf{Future Work}
In our experiment, Ideal-LLM has demonstrated superior performance over several previous works. However, there remains a gap when compared to some SOTA methods. For instance, in the AST task, our results still need to catch up to those achieved by SeamlessM4T~\cite{23seamlessm4t} and the method proposed in~\cite{24s2tt}. We attribute this discrepancy primarily to the limited amount of supervised data available for Ideal-LLM, which cannot match the scale and efficacy of SOTA approaches.
In the future, we plan to leverage larger-scale multilingual datasets to enhance the training of our model.

\section{Conclusion}
\label{sec:conclusion}
In this study, we propose a novel approach, Ideal-LLM, to multilingual speech-to-text tasks by integrating dual encoders with a language-adapted LLM. Our method leverages the complementary strengths of the Whisper and MMS encoders, optimizing their fusion through a CTC loss function and Weight Selector mechanism. The experimental results demonstrate significant improvements in WER across multiple languages, with a 32.6\% relative decrease in average WER compared to the baseline. Additionally, our ablation studies underscore the critical role of the dual-encoder structure and the CTC loss in enhancing performance. This work highlights the potential of combining dual multilingual speech encoders with LLMs to achieve robust and adaptive multilingual speech-to-text, paving the way for more effective and inclusive language processing technologies. We will utilize larger-scale data to train our model in the future.

\vfill\pagebreak

\clearpage

\bibliographystyle{IEEEbib}
\bibliography{refs}

\end{document}